\documentclass[final,3p,times]{elsarticle}

\usepackage{amssymb}

\makeatletter
\def\Hline{%
\noalign{\ifnum0=`}\fi\hrule \@height 1.5pt \futurelet
\reserved@a\@xhline}
\makeatother
\journal{Journal of Computational Physics}

\begin{document}

\begin{frontmatter}

%% Title, authors and addresses

%% use the tnoteref command within \title for footnotes;
%% use the tnotetext command for theassociated footnote;
%% use the fnref command within \author or \address for footnotes;
%% use the fntext command for theassociated footnote;
%% use the corref command within \author for corresponding author footnotes;
%% use the cortext command for theassociated footnote;
%% use the ead command for the email address,
%% and the form \ead[url] for the home page:
%% \title{Title\tnoteref{label1}}
%% \tnotetext[label1]{}
%% \author{Name\corref{cor1}\fnref{label2}}
%% \ead{email address}
%% \ead[url]{home page}
%% \fntext[label2]{}
%% \cortext[cor1]{}
%% \address{Address\fnref{label3}}
%% \fntext[label3]{}

\title{A flux-splitting method for hyperbolic-equation system of magnetized electron fluids in quasi-neutral plasmas}

%% use optional labels to link authors explicitly to addresses:
%% \author[label1,label2]{}
%% \address[label1]{}
%% \address[label2]{}

\author[label1]{Rei Kawashima\corref{cor1}}

\cortext[cor1]{Corresponding author. Tel.: +81 3 5841 6559; fax: +81 3 5841 6559.}

\address[label1]{Department of Aeronautics and Astronautics, The University of Tokyo, 7-3-1 Hongo, Bunkyo-ku, Tokyo 113-8656, Japan}

\author[label1]{Kimiya Komurasaki}

\author[label1]{Tony Sch\"{o}nherr}

\begin{abstract}
   A flux-splitting method is proposed for the hyperbolic-equation system (HES) of magnetized electron fluids in quasi-neutral plasmas.
   The numerical fluxes are split into four categories, which are computed by using an upwind method which incorporates a flux-vector splitting (FVS) and advection upstream splitting method (AUSM).
   The method is applied to a test calculation condition of uniformly distributed and angled magnetic lines of force.
   All of the pseudo-time advancement terms converge monotonically and the conservation laws are strictly satisfied in the steady state.
   The calculation results are compared with those computed by using the elliptic-parabolic-equation system (EPES) approach using a magnetic-field-aligned mesh (MFAM).
   Both qualitative and quantitative comparisons yield good agreements of results, indicating that the HES approach with the flux-splitting method attains a high computational accuracy.
\end{abstract}

\begin{keyword}
   plasma simulation \sep electron fluid \sep hyperbolic equation \sep upwind method
%% keywords here, in the form: keyword \sep keyword

%% PACS codes here, in the form: \PACS code \sep code

%% MSC codes here, in the form: \MSC code \sep code
%% or \MSC[2008] code \sep code (2000 is the default)

\end{keyword}
\end{frontmatter}

%% \linenumbers

%% main text

%%%%%%%%%%%%%%%%%%%%%%%%%%%%%%%%%%%%%%%%%%%%%%%%%%%%%%%%%%%%%
\section{Introduction}
   Quasi-neutral plasma flows appear in many practical applications such as space propulsion, astrophysics, and nuclear fusion \cite{Parra,kajimura2010hybrid,GRL:GRL8860,PLA:602472}.
   In plasma simulations of these applications, the model of lightweight electrons is usually considered separately from the models of ions and neutral particles.
   When considering electron fluids in quasi-neutral plasmas, the space potential is solved by using the electron conservation equations, rather than by using Gauss's law \cite{chen1984introduction}.
   Because of the electrical neutrality, the electron number density distribution is given by the ion number density distribution.
   Therefore, the electron velocity, electron temperature, and space potential are calculated by using the conservation equations for mass, momenta, and energy of electrons.

   The conventional approaches utilize an elliptic equation and a parabolic equation for solving for the space potential and electron temperature \cite{Komurasaki:1995fk,Rognlien1992347}.
   In what follows, this approach will be referred to as the elliptic-parabolic-equation system (EPES) approach.
   However, in the case of magnetized electrons, the EPES becomes an anisotropic diffusion problem, owing to magnetic confinement.
   Computation of this system becomes difficult owing to: 1) the anisotropy stemming from the large difference between diffusion coefficients in different directions and 2) the instability caused by cross-diffusion terms.
   The cross-diffusion terms are especially difficult to handle because they cause the failure of the diagonal dominance of the coefficient matrix.
   One approach toward avoiding the issue of cross-diffusion terms is to utilize a magnetic-field-aligned mesh (MFAM) \cite{Mikellides}.
   By precisely aligning the computational mesh with the magnetic lines of force, the cross-diffusion terms can be neglected because they stem from the angle between the magnetic lines of force and the mesh.
   However, using an MFAM makes it impossible to use a structured mesh for the body-fitted coordinate system and complicates the evaluation of fluxes on the mesh boundaries.
   Furthermore, once the magnetic field induced by the plasma current is solved, one needs to reconstruct the mesh with varying magnetic lines of force.
   Thus, a practical application of the MFAM is associated with cumbersome coding and implementation steps.

   Recently, a novel approach to solving anisotropic diffusion equations has been proposed, which uses a hyperbolic-equation system (HES) \cite{Kawashima201559}.
   The key idea in this approach is to construct a hyperbolic system by introducing pseudo-time advancement terms.
   This approach avoids the aforementioned difficulties related to the anisotropy and cross-diffusion terms.
   It was confirmed that an anisotropic diffusion problem of space potential was robustly computed by using the HES approach.
   It was proved that the advantage of the HES approach compared with the MFAM-based approach was that it could use a simple structured mesh without increasing the computational cost.

   Although the HES approach demonstrated advantages for calculating magnetized electron flows, there remain two issues constraining the applicability of this approach.
   First, the HES approach was proposed only for mass and momentum conservation equations.
   To simulate plasma devices utilizing the heating of electrons for the plasma generation, the HES must include the energy conservation equation for deriving the electron temperature.
   Another issue is the presence of large numerical viscosity.
   It was reported that the HES approach had a large numerical viscosity arising from the discretization of cross-diffusion terms \cite{Kawashima201559}.
   In light of these issues, the purpose of this paper was to extend the HES to include all conservation laws, and to find a robust method for computing the system of equations.
   In addition, a high-order spatial accuracy method was used with the HES approach for reducing the numerical viscosity.
   As a criterion of computational accuracy, we checked whether the HES approach achieves the same level of computational accuracy as the MFAM-based EPES approach.

\section{Hyperbolic system of conservation laws for electron fluid}
   \subsection{Hyperbolic-system formulation of energy conservation equation}
   The fundamental equations are the two-dimensional equations of electron mass, momentum, and energy conservation in quasi-neutral plasma. 
   Assuming quasi-neutrality, the electron number density is equal to the ion number density.
   Thus, the electron number density is treated as a given distribution. 
   Also, the inertia of the electron fluid is neglected in the momentum conservation equation, because of frequent collisions.
   The detailed processes to derive the fundamental equations for the mass and momentum conservations can be found in Ref. \cite{Kawashima201559}.
   Thus, this section focuses on the equations for the energy conservation.

   Energy conservation is formulated as follows:
   \begin{equation}
   	\frac{\partial}{\partial t}\left(\frac{3}{2}en_{\rm e}T_{\rm e}\right)
   	+\nabla \cdot \left(\frac{5}{2}en_{\rm e}T_{\rm e}\vec{u}_{\rm e}-\frac{5}{2}en_{\rm e}T_{\rm e}\left[\mu\right]\nabla T_{\rm e}\right)
   	 =en_{\rm e}\vec{u}_{\rm e}\cdot \nabla \phi-\alpha e\varepsilon_{{\rm ion}}n_{\rm e}\nu_{{\rm ion}},
   	\label{eq:energy}
   \end{equation}
   where $e$, $n_{\rm e}$, $u_{\rm e}$, $T_{\rm e}$, $\phi$, $\nu_{\rm ion}$, and $\varepsilon_{\rm ion}$ are the elemental charge, electron number density, electron velocity, electron temperature, space potential, ionization collision frequency, and first ionization energy of the gas species used for the plasma generation, respectively. 
   $\alpha$ is a coefficient to handle ionization, excitation, and radiation with a single term, and it is experimentally determined as a function of the electron temperature \cite{Dugan_Sovie_1967}. 
   It is assumed that the conservation of energy also achieves a steady state on the ion time scale.
   The electron mobility tensor $\left[\mu\right]$ on a computational mesh is derived as follows: 
   \begin{equation}
      \left[\mu\right]=\left[
   	\begin{array}{cc}
   	\mu_{\rm x}  & \mu_{\rm c}    \\
   	\mu_{\rm c}  & \mu_{\rm y}
	   \end{array}
   	\right]=\Theta^{-1}\left[\mu\right]_{\rm {mag}}\Theta,
	\end{equation}
   where
	\begin{equation}
   	   \left[\mu\right]_{\rm {mag}}= \left[
	   \begin{array}{cccc}
   	\mu_{||}  & \    \\
	   \  & \mu_{\perp}
   	\end{array}
	   \right]= \left[
   	\begin{array}{cccc}
	   \frac{e}{m_{\rm e}\nu_{\rm {col}}}  & \    \\
   	\  & \frac{\mu_{||}}{1+\left(\mu_{||}\left|B \right|\right)^2}
	   \end{array}
   	\right],\hspace{20pt}
	   \Theta=\left[
   		\begin{array}{cccc}
		   	\cos\theta  & -\sin\theta    \\
   			\sin\theta  & \cos\theta
	   	\end{array}
   	\right].
   	\label{eq:18}
   \end{equation}
   Here, $m_{\rm e}$, $\nu_{\rm col}$, and $B$ are electron mass, total collision frequency, and magnetic flux density, respectively.
   The subscripts $||$ and $\perp$ denote parallel and perpendicular directions of magnetic lines of force, respectively.
   $\Theta$ is the rotation matrix, with the angle between the magnetic lines of force and the computational mesh.

   In quasi-neutral plasmas, the plasma approximation is assumed and the space potential is calculated from the conservation equations for the electrons~\cite{chen1984introduction}.
   In conventional approaches, the mass and momentum conservation equations are integrated into an elliptic equation~\cite{Komurasaki:1995fk,Rognlien1992347}. 
   However, this equation becomes an anisotropic diffusion equation, and it is difficult to maintain stability while computing this equation because the cross-diffusion terms cause failure of the diagonal dominance of the coefficient matrix \cite{Kawashima201559}.
   Alternatively, the HES approach using pseudo-time advancement terms is considered.
   The HES which consists of the mass and momentum conservation equations has been proposed in Ref. \cite{Kawashima201559}.

   In addition to the mass and momentum conservation equations, the energy conservation equation is modified to hyperbolic equations to avoid cross-diffusion terms.
   First of all, the negative potential is defined for the negatively charged electrons as follows:
   \begin{equation}
   	\phi_{\rm n}=-\phi.
   	\label{eq:negativepot}
   \end{equation}
   The Joule heating is converted into the potential flux and potential gain by the electron production, by using the equation of continuity, as follows:
   \begin{equation}
   	en_{\rm e}\vec{u}_{\rm e}\cdot \nabla \phi
   	=-en_{\rm e}\vec{u}_{\rm e}\cdot \nabla \phi_{\rm n}
   	=-\nabla \cdot \left(en_{\rm e}\phi_{\rm n}\vec{u}_{\rm e}\right)+en_{\rm e}\phi_{\rm n}\nu_{\rm ion}.
   \label{eq:potflux}
   \end{equation}
   To express Eq. (\ref{eq:energy}) with hyperbolic equations, new variables, $\vec{g}=-\left[\mu\right]\nabla T_e$, are introduced.
   Then, Eq. (\ref{eq:energy}) can be rewritten as follows:
   \begin{equation}
   	\frac{\partial}{\partial t}\left(\frac{3}{2}n_{\rm e}T_{\rm e}\right)
   	+\nabla \cdot \left(\frac{5}{2}n_{\rm e}T_{\rm e}\vec{u}_{\rm e}+n_{\rm e}\phi_{\rm n}\vec{u}_{\rm e}
   	+\frac{5}{2}n_{\rm e}T_{\rm e}\vec{g}\right)
   	=\left(\phi_{\rm n}-\alpha \varepsilon_{\rm ion}\right)n_{\rm e}\nu_{{\rm ion}}.
   	\label{eq:hes3}
   \end{equation}
   The newly introduced variables $\vec{g}$ have the dimensions of velocity, and they are solved by using the pseudo-time advancement terms, as follows:
   \begin{equation}
   	\frac{1}{\nu_{\rm col}}\left[
   	\begin{array}{cc}
   		c_{\rm x} & \  \\
   		\  & c_{\rm y}
   	\end{array}
   	\right]^{-1}\frac{\partial \vec{g}}{\partial t}+[\mu]\nabla T_{\rm e}=-\vec{g},
   	\label{eq:hes4}
   \end{equation}
   where $c_{\rm x}$ and $c_{\rm y}$ are dimensionless acceleration coefficients.
   Eqs. (\ref{eq:hes3}) and (\ref{eq:hes4}) are the hyperbolic equations for the energy conservation.

   Eventually, the HES consists of the hyperbolic equations for the mass, momentum, and energy conservations. Upwind schemes based on the approximate Riemann solver can be constructed by using the Jacobian matrices of the hyperbolic system. This approach, using a first-order system, has been originally proposed by Nishikawa for future application to the Navier-Stokes equation \cite{Nishikawa2007315}. 
   This approach is assumed to be beneficial for the computation of magnetized electrons because it avoids the difficulties associated with cross-diffusion terms.

   \subsection{Nondimensionalization}
   For the analysis, the HES is expressed in a nondimensional form.
   First of all, the electron mobility tensor is normalized by the electron mobility in the direction tangential to the magnetic lines of force, as follows:
   \begin{equation}
      \tilde{\left[\mu\right]}=\left[
   \begin{array}{cccc}
   \tilde{\mu}_{\rm x}  & \tilde{\mu}_{\rm c}    \\
   \tilde{\mu}_{\rm c}  & \tilde{\mu}_{\rm y}
   \end{array}
   \right]=\frac{1}{\mu_{||}}\left[\mu\right]
   =\frac{m_{\rm e}\nu_{\rm {col}}}{e}\left[\mu\right],
   \end{equation}
   where the quantities with tilde denote dimensionless forms.
   The definitions of dimensionless forms for the main physical quantities can be found in Ref. \cite{Kawashima201559}.
   The dimensionless value of $\vec{g}$ is defined as follows:
   \begin{equation}
      \vec{\tilde{g}}=\frac{\vec{g}}{c_{\rm s}^*}=\frac{\vec{g}}{\sqrt{\frac{\gamma eT_{\rm e}^*}{m_{\rm e}}}},
   \end{equation}
   where $c_{\rm s}$, $\gamma$, and $T_{\rm e}^*$ are the electron acoustic velocity, specific heat ratio, and representative electron temperature, respectively. 
   By using the dimensionless forms, a dimensionless equation system can be constructed. 
   Furthermore,
   \begin{equation}
   	\tilde{n}_{\rm e}=1,\hspace{20pt}
   	\tilde{\nu}_{{\rm col}}=1,\hspace{20pt}
   	\tilde{\nu}_{\rm {ion}}=0,
   	\label{eq:assumption}
   \end{equation}
   are assumed for simplicity, throughout the calculation region.

   Optimal choice of the acceleration parameters improves the numerical problems's conditions and accelerates the convergence to a steady-state.
   In Ref. \cite{Kawashima201559}, the authors proposed the acceleration coefficients for avoiding the stiffness stemming from magnetic confinements.
   Following that proposed scheme, the acceleration coefficients in Eq. (\ref{eq:hes4}) were chosen here as follows:
   \begin{equation}
      c_{\rm x}=\frac{\sqrt{2\gamma}}{\tilde{\mu}_{\rm x}},\hspace{20pt}
      c_{\rm y}=\frac{\sqrt{2\gamma}}{\tilde{\mu}_{\rm y}}.
   \end{equation}
   Eventually, the simplified system of dimensionless equations with optimized acceleration coefficients is expressed as follows:
   \begin{equation}
      \frac{\partial \tilde{\phi}_{\rm n}}{\partial \tilde{t}}
      +\tilde{\nabla} \cdot \vec{\tilde{u}}_{\rm e}
      =0,
      \label{eq:hes1_nod}
   \end{equation}
   \begin{equation}
   	\frac{\partial \tilde{\vec{u}_{\rm e}}}{\partial \tilde{t}}
   	+ \left[r\right]\tilde{\nabla} \tilde{\phi}_{\rm n}
   	+ \left[r\right]\tilde{\nabla} \tilde{T}_{\rm e}
   	=-\sqrt{2\gamma}\left[
   	\begin{array}{cccc}
   		\frac{1}{\tilde{\mu}_{\rm x}}  & \ \\
   		\  & \frac{1}{\tilde{\mu}_{\rm y}}
   	\end{array}
   	\right]\vec{\tilde{u}}_{\rm e},
   	\label{eq:hes2_nod}
   \end{equation}
   \begin{equation}
   	\frac{\partial}{\partial \tilde{t}}
   	\left(\frac{3}{2}\tilde{T}_{\rm e}\right)
   	+\tilde{\nabla}\cdot\left(
   	\frac{5}{2}\tilde{T}_{\rm e}\vec{\tilde{u}}_{\rm e}
   	+\tilde{\phi}_{\rm n}\vec{\tilde{u}}_{\rm e}
   	+\frac{5}{2}\tilde{T}_{\rm e}\vec{\tilde{g}}
   	\right)
   	 = 0,
   	\label{eq:hes3_nod}
   \end{equation}
   \begin{equation}
   	\frac{1}{\tilde{\nu}_{{\rm col}}}
   	\frac{\partial \tilde{\vec{g}}}{\partial \tilde{t}}
   	+\left[r\right]\tilde{\nabla} \tilde{T}_{\rm e}
   	=-\sqrt{2\gamma}\left[
   	\begin{array}{cccc}
   		\frac{1}{\tilde{\mu}_{\rm x}}  & \ \\
   		\  & \frac{1}{\tilde{\mu}_{\rm y}}
   	\end{array}
   	\right]\tilde{\vec{g}},
   	\label{eq:hes4_nod}
   \end{equation}
   \begin{equation}
   	\left[r\right]=\left[
   	\begin{array}{cccc}
   		1  & \frac{\tilde{\mu}_{\rm c}}{\tilde{\mu}_{\rm x}}    \\
   		\frac{\tilde{\mu}_{\rm c}}{\tilde{\mu}_{\rm y}}  & 1
   	\end{array}
   	\right].
   	\label{eq:hes5_nod}
   \end{equation}

   %%%%%%%%%%
   \subsection{Elliptic equation and parabolic equation in MFAM}
   In the conventional approach, the elliptic equation that is derived by integrating the mass conservation equation and momentum conservation equation, is used.
   In the coordinate system aligned with magnetic lines of force, this elliptic equation does not include the cross-diffusion terms.
   Using the assumptions in Eq. (\ref{eq:assumption}), this equation can be written as follows:
   \begin{equation}
   	\tilde{\nabla}_{\rm m}\cdot\left(\left[r\right]_{\rm m}
   	\tilde{\nabla}_{\rm m}\left(\tilde{\phi}_{\rm n}+\tilde{T}_{\rm e}\right)\right)=0,
   	\label{eq:mfam1}
   \end{equation}
   \begin{equation}
   	\left[r\right]_{\rm m}=\left[\begin{array}{cc}
   		1 &  \ \\
   		\ &  \frac{\tilde{\mu}_{\perp}}{\tilde{\mu}_{||}} \\
   	\end{array}\right]
   	,\hspace{20pt}\tilde{\nabla}_{\rm m}=\left(\frac{\partial}{\partial \xi},\ \  \ \frac{\partial}{\partial \eta}\right)^T,
   	\label{eq:mfam4}
   \end{equation}
   where $\xi$ and $\eta$ are, respectively, the tangential and orthogonal directions in the MFAM coordinate system.
   Eq. (\ref{eq:mfam1}) is solved for the space potential, and the electron velocity is given by the momentum conservation equation, as follows:
   \begin{equation}
   	\vec{\tilde{u}}_{\rm e}=-\left[r\right]_{\rm m}\tilde{\nabla}_{\rm m}\left(\tilde{\phi}_{\rm n}+\tilde{T}_{\rm e}\right).
   	\label{eq:mfam2}
   \end{equation}
   The parabolic energy conservation equation for deriving the electron temperature can be written as follows:
   \begin{equation}
   	\frac{\partial}{\partial \tilde{t}}
   	\left(\frac{3}{2}\tilde{T}_{\rm e}\right)+
   	\tilde{\nabla}_{\rm m} \cdot \left(
   			\frac{5}{2}\tilde{T}_{\rm e}\vec{\tilde{u}}_{\rm e}
   			-\frac{5}{2}\tilde{T}_{\rm e}\left[r\right]_{\rm m}
   	\tilde{\nabla}_{\rm m} \tilde{T}_{\rm e}
   			-\tilde{\phi}\vec{\tilde{u}}_{\rm e}
   		\right)
   	= 0.
   	\label{eq:mfam3}
   \end{equation}

%%%%%%%%%%%%%%%%%%%%
\section{An upwind flux-splitting method for the HES}
   \subsection{The flux-splitting method}
   An upwind method should be used for robust calculation of the HES with small numerical viscosity.
   To construct an upwind method based on the approximate Riemann solver, the eigenstructures of the flux Jacobian matrices are needed.
   However, in the case of the electron fluid HES, the flux Jacobian matrices are too complicated for analyzing and determining the eigenstructures.
   This issue also arises when the HES approach is applied to the Navier-Stokes equation \cite{Nishikawa20103989}.
   One approach toward calculating a complicated system is to use an approximate Riemann solver that does not require the eigenstructures.
   For instance, the Lax-Friedrichs scheme is usually used for MHD simulations \cite{Kubota:2009aa,Nishida20093182}.
   However, it is known that the Lax-Friedrichs scheme has a large numerical viscosity \cite{Nessyahu1990408}.
   The HES approach for anisotropic diffusion equations originally has a large numerical viscosity arising from cross-diffusion terms.
   Therefore, the Lax-Friedrichs scheme should not be used to avoid excessive numerical viscosities.

   An alternative approach is to use a flux-splitting method such as the advection upstream splitting method (AUSM).
   The AUSM splits the fluxes into some categories based on the physical similarity, and the upwind direction is determined for each split flux.
   The interactions of variables in different fluxes are not explicitly treated in the Jacobian matrices; hence, an upwind method can be constructed without the eigenstuructures. 
   
\begin{figure}[t]
   	\begin{center}
   		\includegraphics[height=50mm]{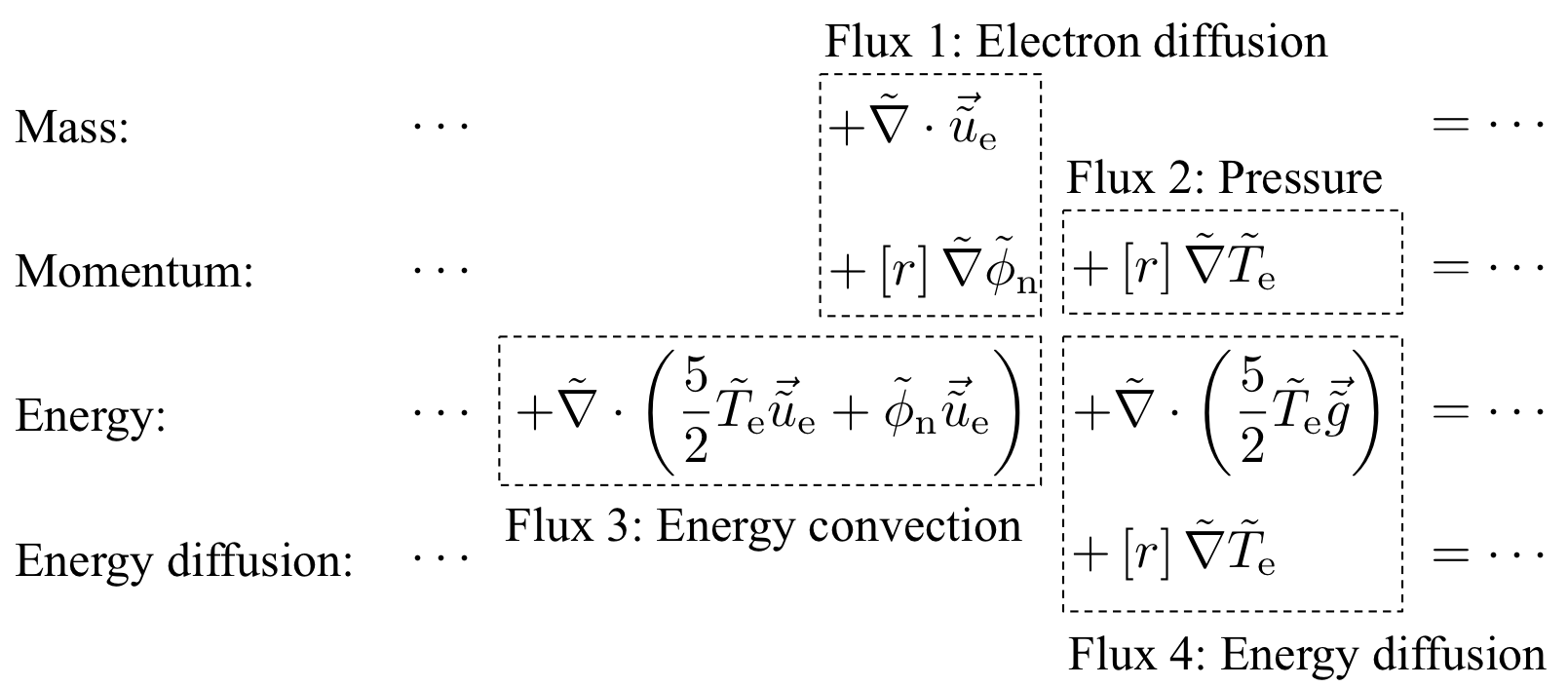}
   	\end{center}
   	\caption{Categorization of numerical fluxes in the HES of Eqs. (\ref{eq:hes1_nod}) - (\ref{eq:hes4_nod}).}
   	\label{fig:category}
   \end{figure}

   For the HES of the electron fluid, the fluxes are split into four categories, as shown in Fig. \ref{fig:category}, based on the physical similarities.
   Flux 1 is the electron diffusion flux induced by potential gradients.
   Flux 2 is the pressure term.
   Flux 3 is the energy convection flux.
   Flux 4 is the energy diffusion flux induced by the gradient of electron temperature.
   The separation of the pressure term from the system is based on the idea of the AUSM.
	Also, the decoupling between the convection and diffusion terms is found in Nishikawa's first-order system approach, which treats inviscid and viscous terms separately \cite{Nishikawa:2011aa}.
	Finally, the diffusion phenomena are split into two fluxes: electron diffusion, which involves space potential and electron velocity, and energy diffusion, which involves electron temperature and energy diffusion speed.
	
	It is noted that modifications can be considered for the flux categorization, depending on problems. 
	The AUSM for the Euler equation also has variations regarding categorization methods of numerical fluxes \cite{Zha1993}. 
	It has been reported that there exists cases to which some of the flux-categorization methods in AUSM are not applicable \cite{Toro20121}. 
	So far, such exceptional cases have not been found for the proposed flux-categorization method.
	
   \subsection{An upwind method incorporating FVS and AUSM}
\begin{figure}[t]
   	\begin{minipage}{0.5\hsize}
   		\begin{center}
   			\includegraphics[height=50mm]{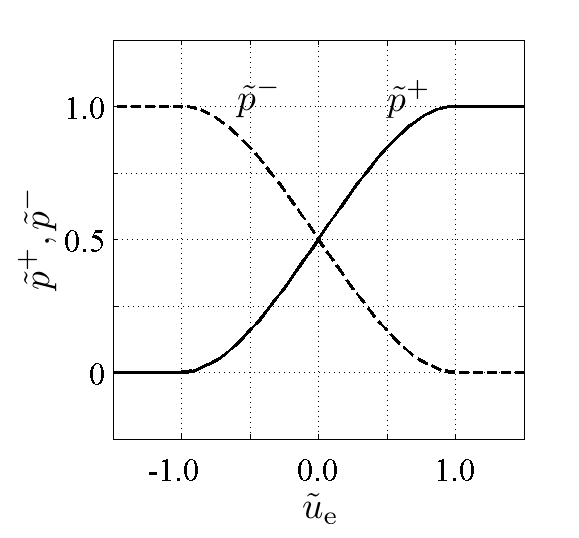}
   		\end{center}
   	\end{minipage}
   	\begin{minipage}{0.5\hsize}
   		\begin{center}
   			\includegraphics[height=50mm]{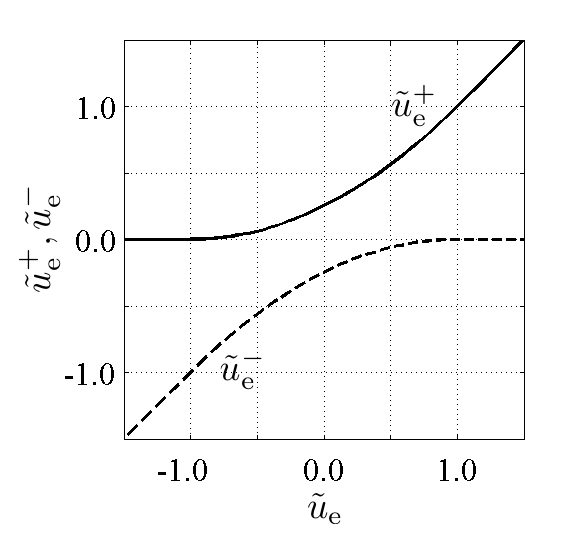}
   		\end{center}
   	\end{minipage}
   	\caption{Left: \(\tilde{p}_x^+\) and \(\tilde{p}_x^-\) as functions of \(\tilde{u}_{{\rm e},x}\). Right: \(\tilde{u}_{{\rm e},x}^+\) and \(\tilde{u}_{{\rm e},x}^-\) as functions of \(\tilde{u}_{{\rm e},x}\). These relationships are equalized in Eq. (\ref{eq:ausm_nod_pres}) and Eq. (\ref{eq:ausm_nod_conv}), respectively.}
   	\label{fig:ausm_nod}
   \end{figure}

   To determine the upwind directions of the fluxes in the four categories, the flux-vector splitting (FVS) \cite{Steger1981263} and AUSM \cite{Liou1996364} are utilized. The upwind directions of the electron diffusion flux (flux 1) and energy diffusion flux (flux 4) are determined by using the FVS.
   The Jacobian matrices and eigenvalues of these fluxes are as simple as follows:
   \begin{equation}
   	J_{\rm x, flux 1}=\left(
   		\begin{array}{ccc}
   			0 & 1 & 0 \   \\
   			1 & 0 & 0 \   \\
   			\frac{\tilde{\mu}_{\rm c}}{\tilde{\mu}_{\rm y}} & 0 & 0
   		\end{array}
   	\right),\hspace{20pt}
   	J_{\rm y, flux 1} = \left(
   		\begin{array}{ccc}
   			0 & 0 & 1 \   \\
   			\frac{\tilde{\mu}_{\rm c}}{\tilde{\mu}_{\rm x}} & 0 & 0 \   \\
   			1 & 0 & 0
   		\end{array}
   	\right),
   	\label{eq:Jacob-1}
   \end{equation}
   \begin{equation}
   	\lambda_{\rm x, flux 1}=0, \pm 1,\hspace{20pt}
   	\lambda_{\rm y, flux 1} = 0, \pm 1,
   	\label{eq:lambda-1}
   \end{equation}
   \begin{equation}
   	J_{\rm x, flux 4}=\left(
   		\begin{array}{ccc}
   			\frac{5}{2}g_x & \frac{5}{2}T_{\rm e} & 0 \   \\
   			1 & 0 & 0 \   \\
   			0 & 0 & 0
   		\end{array}
   	\right),\hspace{20pt}
   	J_{\rm y, flux 4} = \left(
   		\begin{array}{ccc}
   			\frac{5}{2}g_y & 0 & \frac{5}{2}T_{\rm e} \   \\
   			0 & 0 & 0 \   \\
   			1 & 0 & 0
   		\end{array}
   	\right),
   	\label{eq:Jacob-4}
   \end{equation}
   \begin{equation}
   	\lambda_{\rm x, flux 4}=0,
   	\frac{5}{4}g_{\rm x}\pm \sqrt{\frac{25}{16}g_{\rm x}^2+\frac{5}{2}T_{\rm e}},
   	\hspace{20pt}
   	\lambda_{\rm y, flux 4} = 0,
   	\frac{5}{4}g_{\rm y}\pm \sqrt{\frac{25}{16}g_{\rm y}^2+\frac{5}{2}T_{\rm e}}.
   	\label{eq:lambda-4}
   \end{equation}
   It is noted that the signs of the eigenvalues in Eq. (\ref{eq:lambda-1}) and Eq. (\ref{eq:lambda-4}) are fixed, and hence the flux-vector splitting method does not cause instabilities related to the change of eigenvalue signs.
   The upwind splittings of the pressure flux (flux 2) and energy convection flux (flux 3) are determined by using the AUSM.
   The nondimensional electron temperature in the pressure term is split into right- and left-running waves of information, as follows:
   \begin{equation}
   	\tilde{T}_{\rm e}=\tilde{T}_{\rm e}^+ +\tilde{T}_{\rm e}^- = \tilde{p}^+\tilde{T}_{\rm e}+\tilde{p}^-\tilde{T}_{\rm e}.
   	\label{eq:split1}
   \end{equation}
   The coefficients $\tilde{p}^\pm$ are determined based on the evaluation method of pressure term in the AUSM.
   In the x-direction for example, they are given as follows:
   \begin{equation}
   	\tilde{p}_{\rm x}^\pm=\left\{
   	\begin{array}{cc}
   		\frac{1}{2}\left(1\pm {\rm sign}\left(\tilde{u}_{\rm e,x}\right)\right),
   		& {\rm if}\left|\tilde{u}_{\rm e,x}	\right|\geq1,  \\
   		\pm\frac{1}{4}\left(\tilde{u}_{\rm e,x}\pm1\right)^2\left(2\mp \tilde{u}_{\rm e,x}\right),
   		& {\rm otherwise}.
   		\end{array}
   	\right.
   	\label{eq:ausm_nod_pres}
   \end{equation}
   The numerical flux in the energy convection flux is split as follows:
   \begin{equation}
   	\tilde{u}_{\rm e}\left(\frac{5}{2}\tilde{T}_{\rm e}+\tilde{\phi}_{\rm n}\right)
   	=\tilde{u}_{\rm e}^+\left(\frac{5}{2}\tilde{T}_{\rm e}+\tilde{\phi}_{\rm n}\right)+\tilde{u}_{\rm e}^-\left(\frac{5}{2}\tilde{T}_{\rm e}+\tilde{\phi}_{\rm n}\right).
   	\label{eq:split2}
   \end{equation}
   The split electron velocities $\tilde{u}_{\rm e}^\pm$ are associated with the split Mach number in the AUSM.
   In the x-direction for instance, they are formulated as follows:
   \begin{equation}
   	\tilde{u}_{\rm x}^\pm=\left\{
   	\begin{array}{cc}
   		\frac{1}{2}\left(\tilde{u}_{\rm e,x}\pm\left|\tilde{u}_{\rm e,x}\right|\right),
   		& {\rm if}\left|\tilde{u}_{\rm e,x}\right|\geq1,  \\
   		\pm\frac{1}{4}\left(\tilde{u}_{\rm e,x}\pm1\right)^2,
   		& {\rm otherwise}.
   	\end{array}
   	\right.
   	\label{eq:ausm_nod_conv}
   \end{equation}
   In Eqs. (\ref{eq:split1}) and (\ref{eq:split2}), the superscripts ``+" and ``-" correspond to the right- and left-running waves, and the backward and forward discretization is applied, respectively.
   The quantities $\tilde{p}^\pm$ in Eq. (\ref{eq:ausm_nod_pres}) and $\tilde{u}_{\rm e}^\pm$ in Eq. (\ref{eq:ausm_nod_conv}) are shown in Fig. \ref{fig:ausm_nod} as functions of $\tilde{u}_{\rm e}$.
   The polynomial approximations used in these quantities avoid non-differentiable points.

%%%%%%%%%%%%%%%%%%%%
\section{Calculation conditions and numerical methods}
   \subsection{The calculation conditions}
\begin{figure}[t]
		\begin{center}
			\includegraphics[width=90mm]{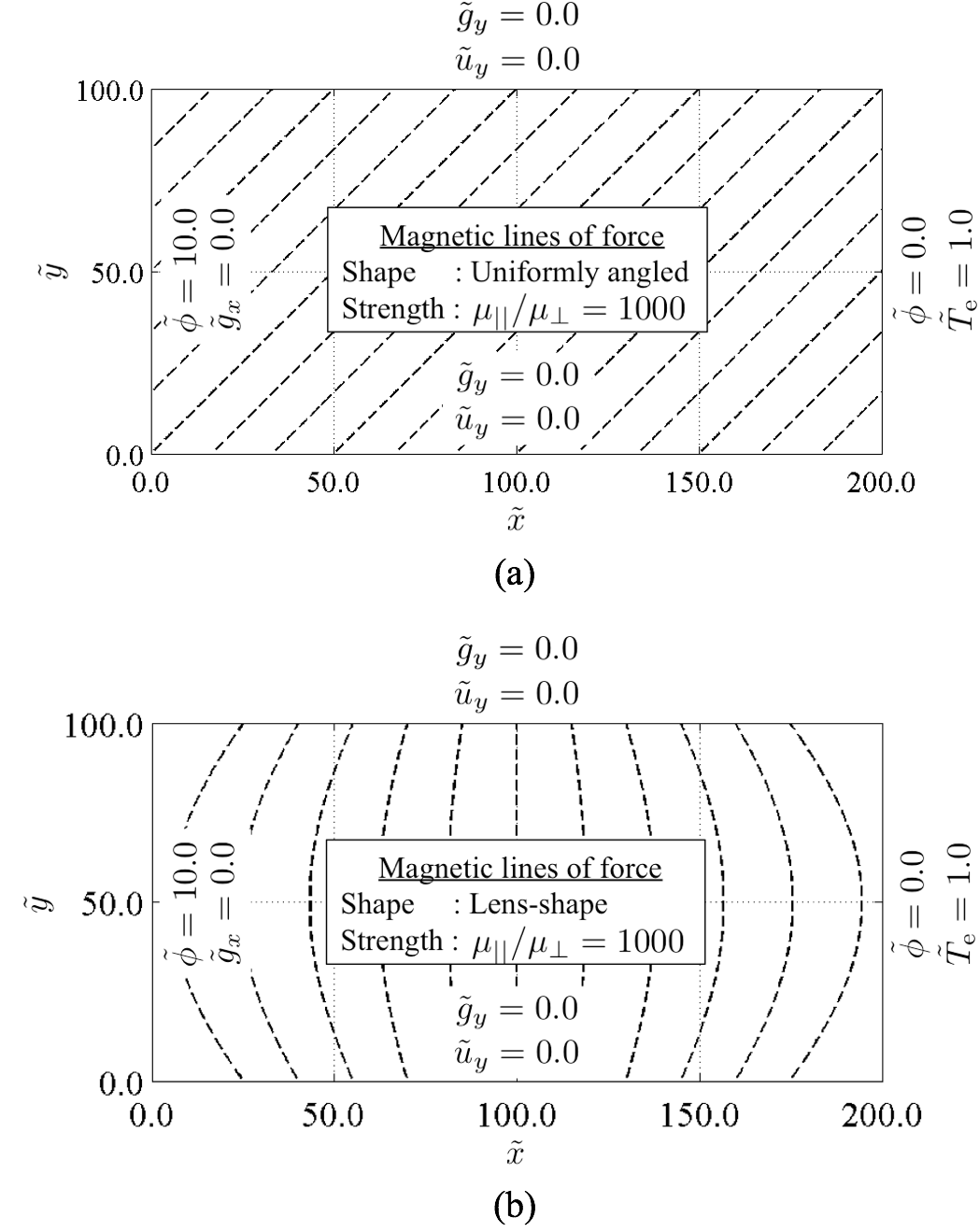}
		\end{center}
		\caption{Two-dimensional calculation conditions of the test calculation. 
		(a) The magnetic lines of force are uniformly distributed and at 45$^\circ$ relative to the horizontal line.
		(b) Lens-shaped magnetic lines of force are applied where the angles of the lines relative to the vertical line are within $\pm$45$^\circ$.
		For both cases the confinement strength of $\mu_{||}/\mu_{\perp}=1000$ is assumed. 
		All boundary conditions are defined by Dirichlet condition. 
		The quantities $\tilde{\phi}$ are defined on the left and right boundaries. 
		The zero-electron-flux condition is used for the top and bottom boundaries. 
		$\tilde{T}_{\rm e}$ is defined on the right boundary.}
		\label{fig:condition_combine}
	\end{figure}

   To verify the applicability of the HES approach with the flux-splitting method, two-dimensional test calculations were conducted.
   The calculation area and boundary conditions of the test calculations are illustrated in Fig. \ref{fig:condition_combine}.
   In Fig. \ref{fig:condition_combine} (a) condition, magnetic lines of force at the angle of 45$^\circ$ relative to the horizontal line were applied throughout the calculation field.
   This calculation condition was used for the qualitative and quantitative comparisons between the results obtained by the HES and MFAM-based EPES approaches.
   The condition of magnetic force lines at 45$^\circ$ yields the strongest effect of cross-diffusion terms if a vertical-horizontal mesh (VHM) is used.
   Thus, if this condition is accurately calculated by using the HES approach, high computational accuracy can be expected for any orientation angle of the magnetic lines of force.
   In Fig. \ref{fig:condition_combine} (b) condition, lens-shaped magnetic lines of force were applied.
   This calculation condition was used to confirm the applicability of the HES approach to a condition with complicated magnetic field geometry.
   In both conditions in Fig. \ref{fig:condition_combine}, the strength of magnetic confinement was uniformly distributed and was $\mu_{||}/\mu_{\perp}=1000$.
   Dirichlet conditions on the nondimensional space potential are defined on the left and right boundaries.
   In addition, the dimensionless electron temperature is defined on the right boundary.
   Zero-flux conditions are assumed for the electron and energy diffusion fluxes on the top and bottom boundaries.

   \subsection{The numerical method used in the HES approach}
   The numerical methods used in the HES approach are summarized in Table \ref{tab:method}.
   A VHM was used in the HES approach for the test calculations.
   The configuration of the VHM can be found in Ref. \cite{Kawashima201559}.
   For comparing the computational accuracy, the HES was discretized by using two different methods: HES-VHM-1ST and HES-VHM-3RD.
   HES-VHM-1ST employs a first-order upwind method, whereas HES-VHM-3RD employs a third-order total variation diminishing monotonic upstream-centered scheme for conservation laws (TVD-MUSCL) technique, coupled with the minmod limiter function.
   In the HES approaches, to make the pseudo-time advancement terms negligibly small, computations must be continued until the steady state is reached.
   In this case, the implicit time advancement should be employed for avoiding the Courant-Friedrichs-Lewy (CFL) restriction.
   In the HES approach, an alternating-direction-implicit symmetric Gauss-Seidel (ADI-SGS) method \cite{Nishida20093182} was implemented with a Courant number of 30.

   \subsection{The numerical method used in the EPES approach}
   The EPES approach was also implemented in the test calculation of Fig. \ref{fig:condition_combine} (a) condition for comparing the computational accuracy. 
   In what follows, this approach is referred to as the EPES-MFAM-2ND approach.
   The numerical methods used in this approach are also summarized in Table \ref{tab:method}.
   The approach is MFAM-based and the configuration of the MFAM can be found in Ref. \cite{Kawashima201559}. 
   The advection terms and diffusion terms in the system were discretized by using a second-order TVD-MUSCL and a second-order central difference, respectively. 
   Eqs. (\ref{eq:mfam1}), (\ref{eq:mfam2}), and (\ref{eq:mfam3}) were computed iteratively to derive the steady-state solution with tolerable residuals.
   Eq. (\ref{eq:mfam1}) was regarded as a boundary value problem for $\tilde{\phi}$, and it was calculated by using a successive over-relaxation (SOR) method. 
   The velocity $\vec{\tilde{u}}_{\rm e}$ was calculated from Eq. (\ref{eq:mfam2}). 
   The parabolic Eq. (\ref{eq:mfam3}) was treated as a time-development problem for $\tilde{T}_{\rm e}$, and the lower-upper symmetric Gauss-Seidel (LU-SGS) method was implemented with a Courant number of 30.

   \begin{table}[t]
   	\caption{Summary of numerical methods for the HES approach and EPES approaches.}
   	\label{tab:method}
   	\begin{center}
   		\begin{tabular}{p{20mm}p{45mm}p{42mm}}
   			\Hline
   			Name & HES-VHM-1ST / -3RD & EPES-MFAM-2ND \\ \hline
   			Equation &  HES &  EPES \\
   			Mesh & VHM  & MFAM \\
   			Convection & First-order upwind / & Second-order TVD-MUSCL  \\
   			\  & Third-order TVD-MUSCL & \  \\
   			Diffusion & -- & Second-order central  \\
   			Iteration & ADI-SGS method & SOR + LU-SGS method \\
   			\Hline
   		\end{tabular}
   	\end{center}
   \end{table}

%\clearpage
   \section{Results and discussion}

   \subsection{Convergence history of normalized differences}
\begin{figure}[t]
   	\begin{center}
   		\includegraphics[width=90mm]{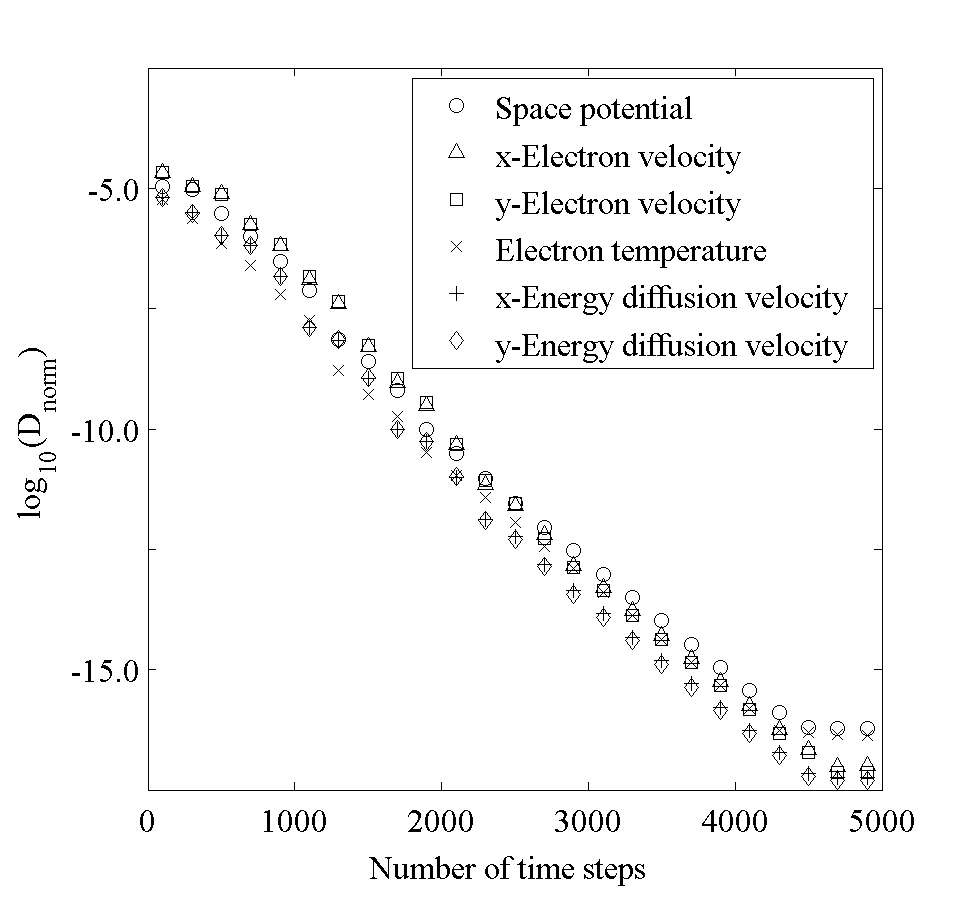}
   	\end{center}
    \caption{Convergence history of the normalized difference for each variable. The condition in Fig. \ref{fig:condition_combine} (a) is solved by using the HES-VHM-3RD method on a 96 $\times$ 48 grid.}
     \label{fig:convergence_45}
   \end{figure}

   First of all, convergences of the pseudo-time advancement terms in the HES were confirmed.
   The pseudo-time advancement terms were artificially introduced into the system. Hence, they must converge to negligibly small values in the steady state. To evaluate the magnitude of the time advancement terms, the normalized difference was defined as follows:

   \begin{equation}
      D_{\rm {norm}}=\sqrt{\frac{1}{N_{\rm {cell}}}\sum^{N_{\rm {cell}}}\left(\frac{|y^{n+1}-y^{n}|^2}{|y^n|^2+\varepsilon}\right)},
      \label{eq:Dnorm}
   \end{equation}
   where \(y\) were the variables (\(\tilde{\phi}, \tilde{u}_{\rm x}, \tilde{u}_{\rm y}, \tilde{T}_{\rm e}, \tilde{g}_{\rm x}, \tilde{g}_{\rm y}\)), and \(\varepsilon\) was a tiny positive value satisfying \(\varepsilon \ll |y|\) for avoiding division by zero.
   The convergence history of $D_{\rm {norm}}$ for each variable is shown in Fig. \ref{fig:convergence_45} for the test problem in Fig. \ref{fig:condition_combine} (a) calculated with the HES-VHM-3RD on a 96 $\times$ 48 grid.
   Normalized differences of all variables decreased monotonically and eventually attained negligibly small values on the order of 10\(^{-16}\) - 10\(^{-18}\).
   This monotone decrease validates the stability of the computation based on the flux-splitting method.
   Further, the attainment of negligibly small normalized differences proves that the conservation laws are strictly satisfied in the steady state.

   %\clearpage
   \subsection{Qualitative comparison: steady-state calculation results}
   The test case in Fig. \ref{fig:condition_combine} (a) was computed by using the HES-VHM-3RD and EPES-MFAM-2ND methods on a 96 $\times$ 48 grid, and the results obtained by using the two approaches were compared.
   A high computational accuracy was expected for the EPES-MFAM-2ND because MFAM-based approach did not suffer from numerical viscosity arising from cross-diffusion terms. 
   Therefore, the purpose of this comparison was to observe whether the HES-VHM-3RD results qualitatively resemble the EPES-MFAM-2ND results.

\begin{figure}[t]
   	\begin{minipage}{0.5\hsize}
   		\begin{center}
   			\includegraphics[width=80mm]{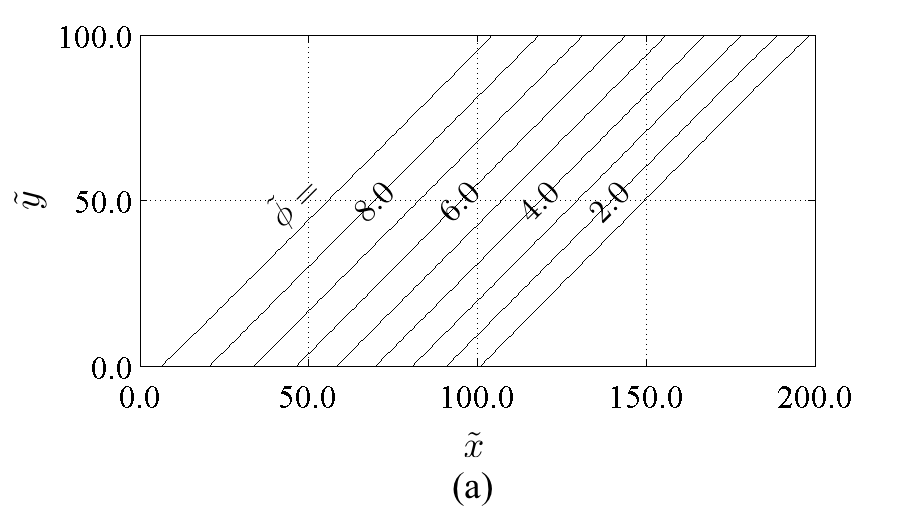}
   		\end{center}
   	\end{minipage}
   	\begin{minipage}{0.5\hsize}
   		\begin{center}
   			\includegraphics[width=80mm]{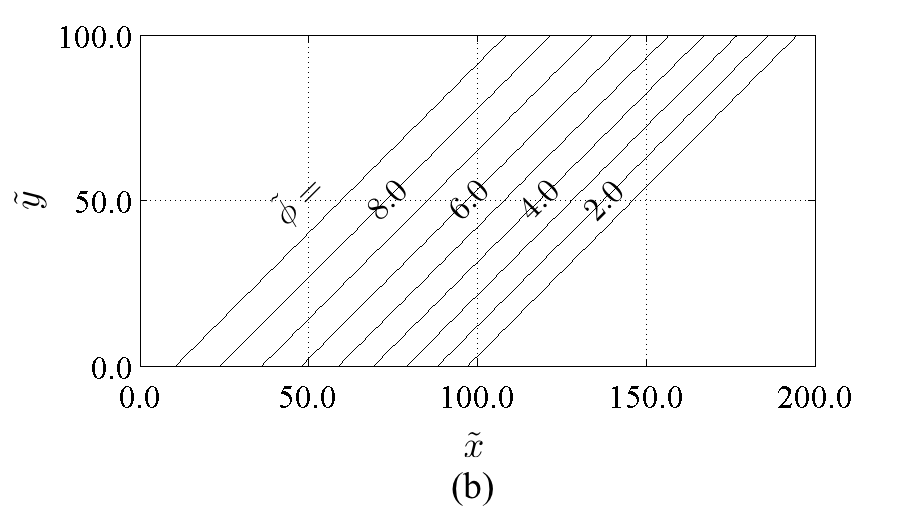}
   		\end{center}
   	\end{minipage}
    \caption{Distribution of dimensionless space potential, calculated by using (a) the HES-VHM-3RD method, and (b) the EPES-MFAM-2ND method.
    	The condition in  Fig. \ref{fig:condition_combine} (a) is computed on a grid of 96 $\times$ 48.}
     \label{fig:phicomp}
   \end{figure}

\begin{figure}[t]
   	\begin{minipage}{0.5\hsize}
   		\begin{center}
   			\includegraphics[width=80mm]{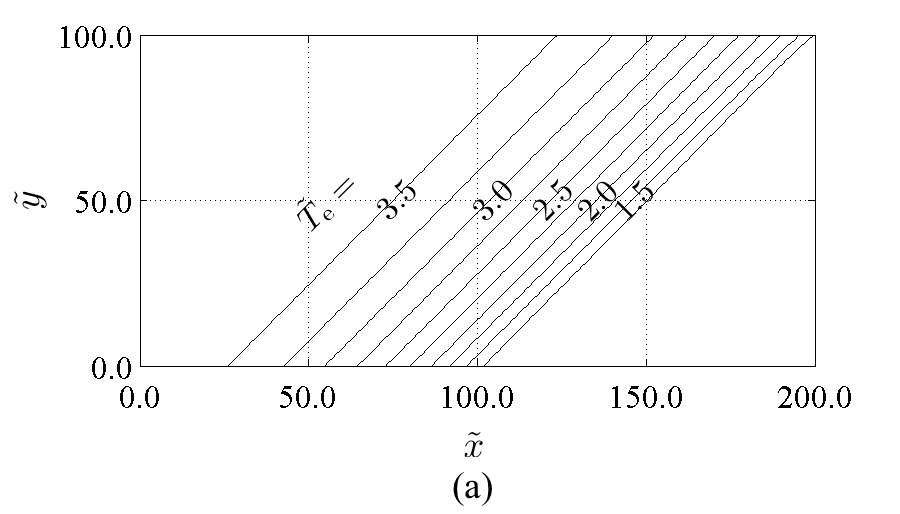}
   		\end{center}
   	\end{minipage}
   	\begin{minipage}{0.5\hsize}
   		\begin{center}
   			\includegraphics[width=80mm]{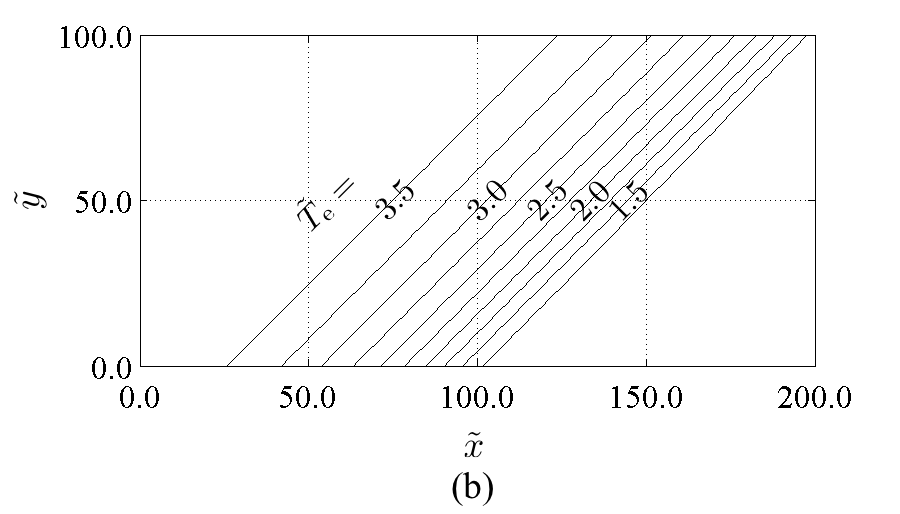}
   		\end{center}
   	\end{minipage}
    \caption{Distribution of dimensionless electron temperature, calculated by using (a) the HES-VHM-3RD method, and (b) the EPES-MFAM-2ND method. 
		The condition in Fig. \ref{fig:condition_combine} (a) is computed on a grid of 96 $\times$ 48.}
		\label{fig:tecomp}
   \end{figure}

\begin{figure}[t]
   	\begin{minipage}{0.5\hsize}
   		\begin{center}
   			\includegraphics[width=80mm]{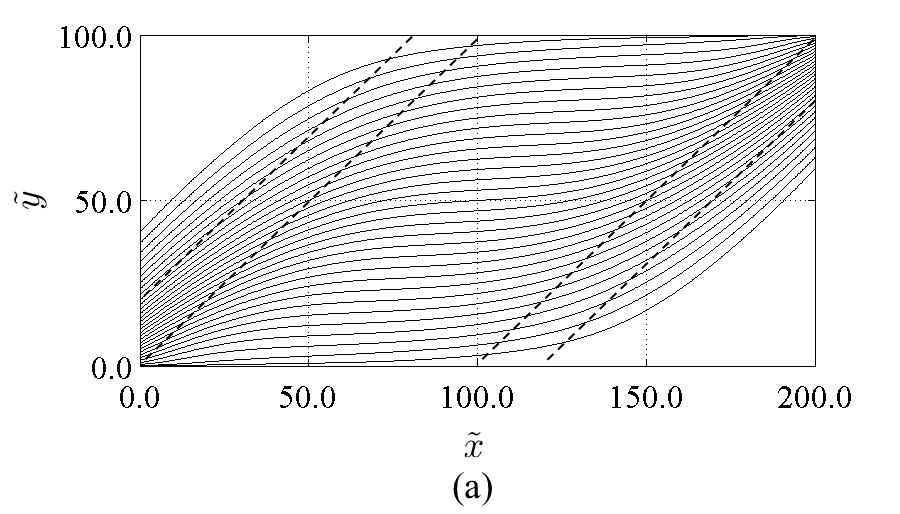}
   		\end{center}
   	\end{minipage}
   	\begin{minipage}{0.5\hsize}
   		\begin{center}
   			\includegraphics[width=80mm]{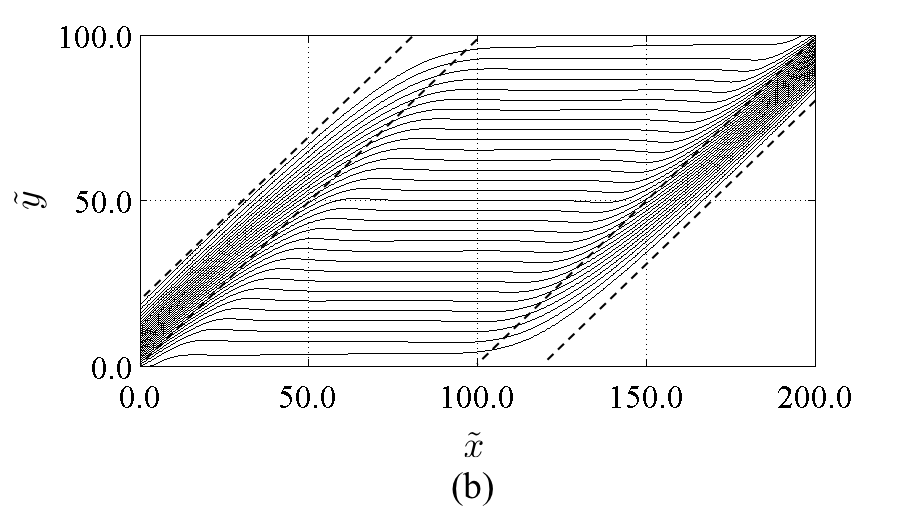}
   		\end{center}
   	\end{minipage}
   		\begin{center}
   			\includegraphics[width=80mm]{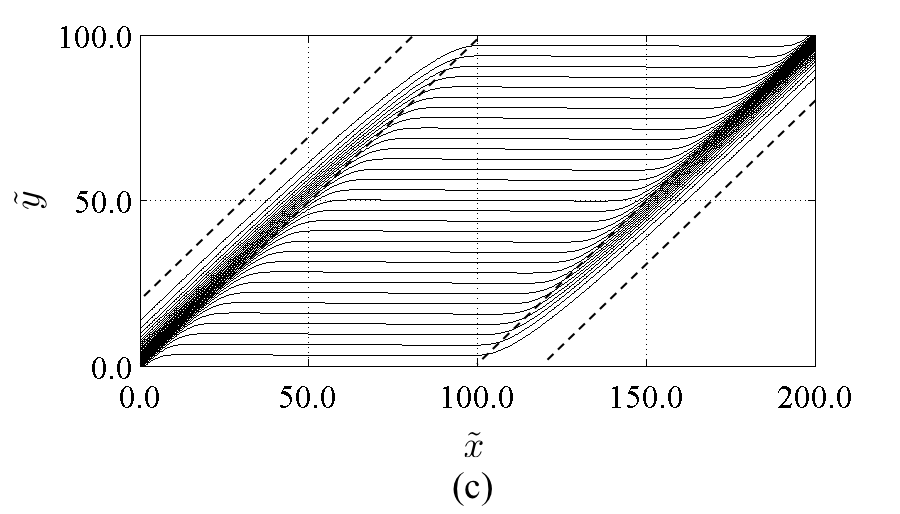}
   		\end{center}
    \caption{Electron streamlines calculated by using (a) the HES-VHM-1ST method, (b) the HES-VHM-3RD method, and (c) the EPES-MFAM-2ND method. 
			The condition in Fig. \ref{fig:condition_combine} (a) is computed on a grid of 96 $\times$ 48. 
			The dashed lines are the reference magnetic lines of force.}
		\label{fig:streamcomp}
   \end{figure}

   The dimensionless space potential and dimensionless electron temperature distribution in the steady state are shown in Figs. \ref{fig:phicomp} and \ref{fig:tecomp}, respectively.
   The space potential and electron temperature distributions, calculated by using the HES-VHM-3RD, were qualitatively similar to those calculated by using the EPES-MFAM-2ND.
   The improvement of computational accuracy obtained by implementing the third-order scheme is clearly seen in the electron streamlines.
   The electron streamlines calculated by using the HES-VHM-1ST, HES-VHM-3RD, and EPES-MFAM-3RD methods are shown in Fig. \ref{fig:streamcomp}.
   In Fig. \ref{fig:streamcomp} (a), the electrons diffuse across the magnetic lines of force owing to the large numerical viscosity, and the effect of magnetic confinement is not reflected accurately.
   However, in Fig. \ref{fig:streamcomp} (b) the numerical viscosity is reduced owing to the third-order scheme, and the streamlines resemble those in Fig. \ref{fig:streamcomp} (c).
   Based on these qualitative comparisons it is concluded that the HES approach with the flux-splitting method attains a high computational accuracy, of the same level as that of the MFAM-based EPES approach.
   
\begin{figure}[t]
   	\begin{minipage}{0.5\hsize}
   		\begin{center}
   			\includegraphics[width=80mm]{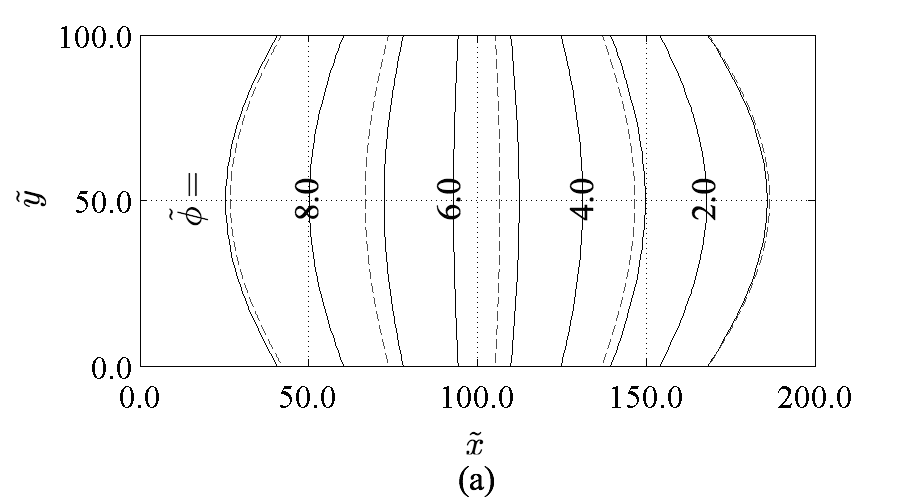}
   		\end{center}
   	\end{minipage}
   	\begin{minipage}{0.5\hsize}
   		\begin{center}
   			\includegraphics[width=80mm]{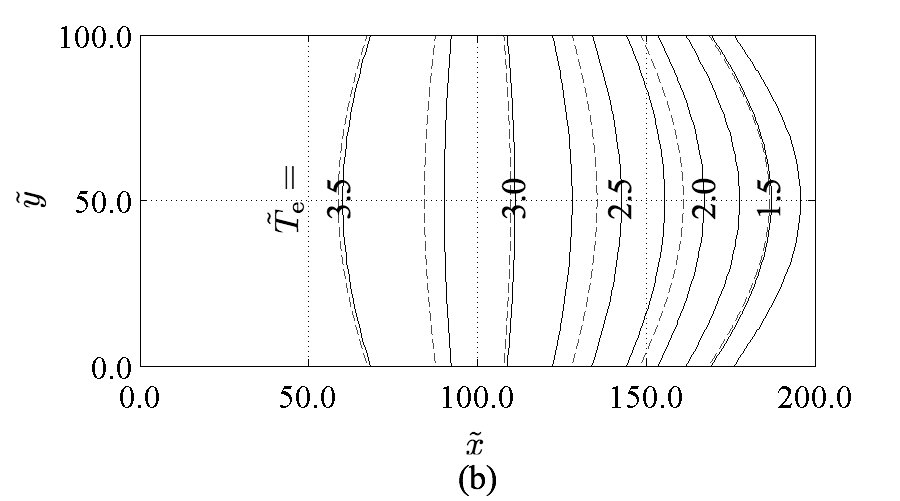}
   		\end{center}
   	\end{minipage}
   		\begin{center}
   			\includegraphics[width=80mm]{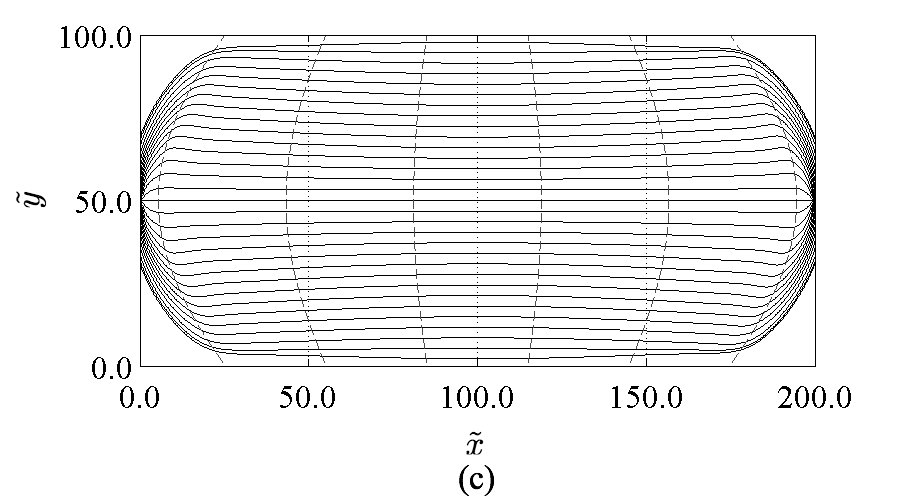}
   		\end{center}
    \caption{Calculation results for Fig. \ref{fig:condition_combine} (b) condition with the HES-VHM-3RD method on a grid of 96 $\times$ 48. 
    	(a) Dimensionless space potential, (b) dimensionless electron temperature, and (c) electron streamlines. 
    	The dashed lines are the reference magnetic lines of force.}
   	\label{fig:curve}
   \end{figure}

   The test case in Fig. \ref{fig:condition_combine} (b) was also computed by using the HES-VHM-3RD method, on a 96 $\times$ 48 grid.
   The purpose of this test calculation is to verify the applicability of the HES approach to a condition with complicated magnetic field configuration.
   In this calculation, the convergence history of each variable showed a monotonic decrease such as the one in Fig. \ref{fig:convergence_45}.
   The calculated distributions of dimensionless space potential, dimensionless electron temperature, and electron streamlines are presented in Fig. \ref{fig:curve}.
   The calculated equipotential lines and isothermal lines are consistent with the magnetic lines of force.
   This indicates that the HES approach accurately computes the condition of complicated magnetic lines of force.
   
%\clearpage
   \subsection{Quantitative comparison: mesh convergence of transverse electron flux and electron heat flux}
\begin{figure}[t]
   	\begin{minipage}{0.5\hsize}
   		\begin{center}
   			\includegraphics[width=70mm]{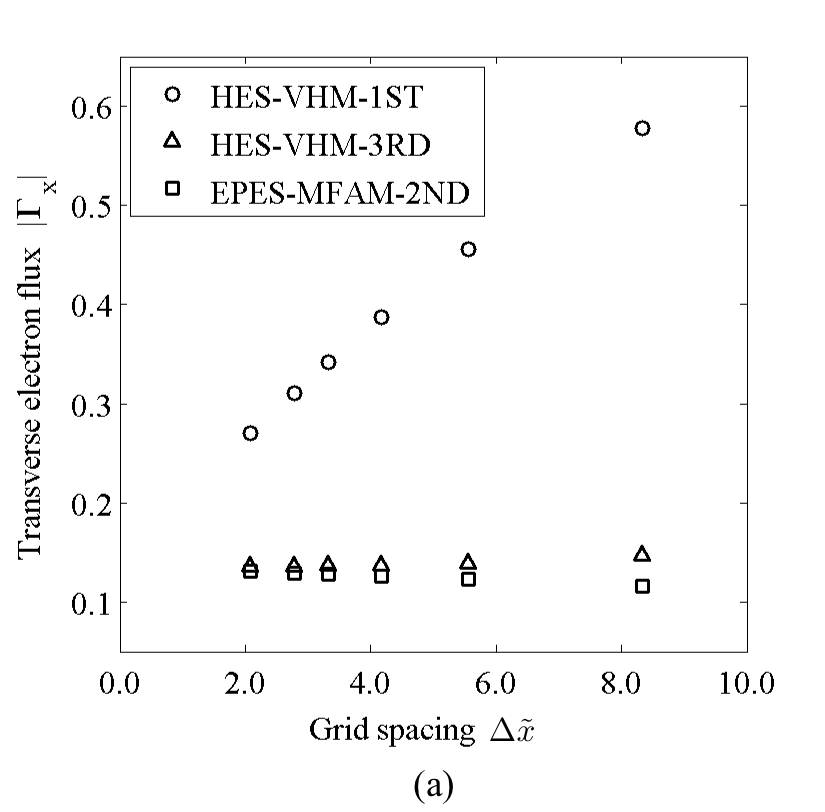}
   		\end{center}
   	\end{minipage}
   	\begin{minipage}{0.5\hsize}
   		\begin{center}
   			\includegraphics[width=70mm]{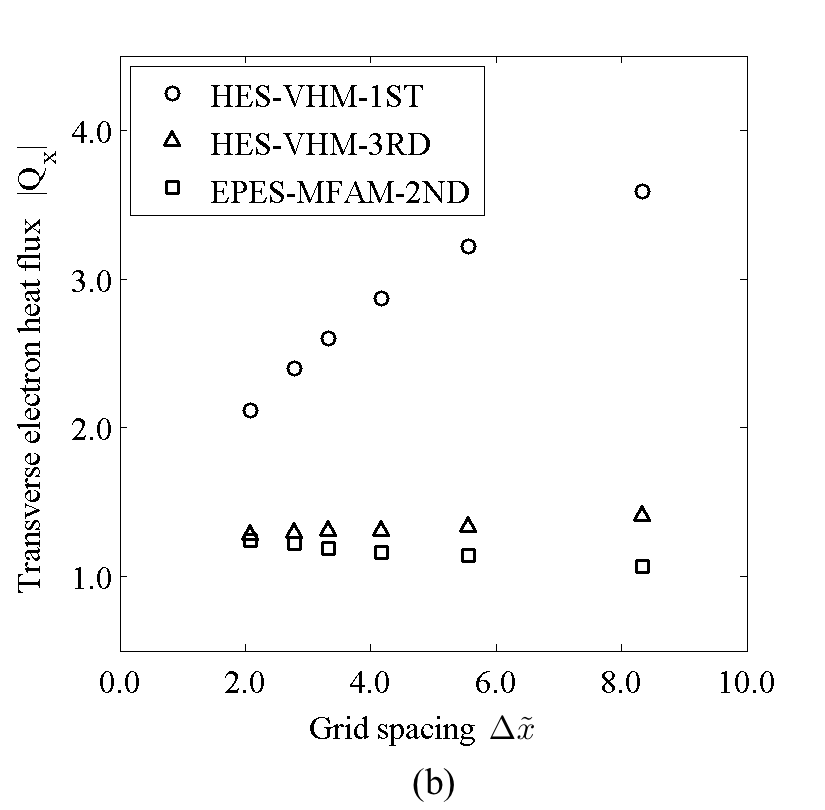}
   		\end{center}
   	\end{minipage}
    \caption{(a) Mesh convergence of transverse electron flux. (b) Mesh convergence of electron heat flux. The condition in Fig. \ref{fig:condition_combine} (a) is solved by using the HES-VHM-1ST, HES-VHM-3RD, and EPES-MFAM-2ND methods and varying the grid spacing.}
     \label{fig:meshconv}
   \end{figure}

   It is difficult to quantitatively analyze the computational accuracy for anisotropic diffusion problems because the analytical solution cannot be derived easily.
   Thus, to quantitatively compare the computational accuracies of the HES and EPES approaches, the mesh convergences of electron and electron heat fluxes traversing the calculation field were evaluated. Here, the transverse electron flux $\Gamma_{\rm x}$ and the transverse electron heat flux $Q_{\rm x}$ were defined as follows:
   \begin{equation}
   	\Gamma_{\rm x}=\int_{\Omega_{\rm L}}\left(
   	\tilde{u}_{\rm e,x}\right)d\tilde{y},\hspace{20pt}
   	Q_{\rm x}=\int_{\Omega_{\rm L}}\left(
   	\frac{5}{2}\tilde{T}_{\rm e}\tilde{u}_{\rm e,x}
   	+\frac{5}{2}\tilde{T}_{\rm e}\tilde{g}_{\rm x}
   	\right)d\tilde{y},
   	\label{eq:transverse}
   \end{equation}
   where $\Omega_{\rm L}$ is the left boundary of the calculation field.
   The mesh convergences of the transverse electron flux and electron heat flux, calculated by using the HES-VHM-1ST, HES-VHM-3RD, and EPES-MFAM-2ND methods, are visualized in Fig. \ref{fig:meshconv}.
   With any grid spacing, the transverse fluxes calculated by using the HES-VHM-1ST method were overestimated owing to a large numerical viscosity of the first-order scheme.
   This implies that the effect of magnetic confinement was not evaluated accurately. However, in the results obtained by using the HES-VHM-3RD method, the numerical viscosity was reduced and both the transverse electron flux and the electron heat flux were in good agreements with the values calculated by using the EPES-MFAM-2ND method in the fine grid system. Even in the coarser grid systems, the deviations of transverse electron and electron heat fluxes from their converged values were small.
   These results imply that a quantitatively high computational accuracy is obtained by using the HES-VHM-3RD method.

\clearpage
   \section{Concluding remarks}
   The HES was constructed for the conservation laws of magnetized electron fluids in quasi-neutral plasmas.
   To construct a robust upwind method based on the approximate Riemann solver for the HES, the flux-splitting method was proposed.
   The fluxes were split into four categories for which the upwind method incorporating the FVS and AUSM was used. The HES approach with the flux-splitting method was tested by using the calculation condition in which the magnetic lines of force were uniformly distributed and at 45$^\circ$ relative to the horizontal direction. The findings are summarized as follows:
   \begin{enumerate}
   	\item All pseudo-time advancement terms converge monotonically to negligibly small values. This fact validates the robustness of the flux-splitting method and that the conservation equations are strictly satisfied.
   	\item The numerical viscosity is drastically reduced by using the high-order scheme, which results in better reflection of the magnetic confinement.
   	\item The calculation results obtained by using the HES approach with the flux-splitting method are consistent with the results obtained by using the MFAM-based EPES approach, in both qualitative and quantitative comparisons.
   	This fact indicates that a high computational accuracy can be obtained by using the HES approach, even with the use of a simple structured mesh.
   \end{enumerate}

%\clearpage
\section*{Acknowledgment}

   This work was supported by the Grant-in-Aid for JSPS Fellows, No. 24-10079.

%\clearpage
\renewcommand{\refname}{Reference}
\bibliographystyle{elsarticle-num}
\bibliography{JCP2nd_reference}

\end{document}